\documentclass[twocolumn]{revtex4}
\usepackage{graphicx}
\usepackage{amsfonts}
\usepackage{amssymb}
\usepackage{array}
 \usepackage{booktabs}

\def\mod{\mbox{ mod }}

\begin{document}
\title[Space representation of stochastic processes with delay]
      {Space representation of stochastic processes with delay}
\author{Silvio R. Dahmen$^{1,2}$, 
Haye Hinrichsen$^1$, 
and Wolfgang Kinzel$^1$}

\address{$^1$ Universit\"at W\"urzburg, Fakult\"at f\"ur Physik und Astronomie,
	 Am Hubland, D-97074 W\"urzburg, Germany}
\address{$^2$ Instituto de F\'{\i}sica da UFRGS,
	Av. Bento Gon{\c c}alves 9500, 91501-970 Porto Alegre, Brazil}

\begin{abstract}
We show that a time series $x_t$ evolving by a \textit{non-local} update rule $x_t = f (x_{t-n},x_{t-k})$ with two different delays $k<n$ can be mapped onto a \textit{local} process in two dimensions with special time-delayed boundary conditions provided that $n$ and $k$ are coprime. For certain stochastic update rules exhibiting a non-equilibrium phase transition this mapping implies that the critical behavior does not depend on the short delay $k$. In these cases, the autocorrelation function of the time series is related
to the critical properties of directed percolation.
\end{abstract}

\parskip 2mm 
\vglue 5mm
\maketitle

\section{Introduction}

Dynamical systems with time-delayed feedback show interesting
phenomena which are relevant to a broad spectrum of research fields,
such as nonlinear dynamics~\cite{strogatz}, laser physics~\cite{wuensche}, neurobiology~\cite{cho}, chaos
control~\cite{boccaletti}, synchronization~\cite{pikovsky,szendro} and
communication~\cite{argyris}. The mathematics of delayed differential
equations is less developed and not as understood as that of ordinary ones~\cite{hale}. Hence the
theoretical analysis of delayed feedback currently attracts a lot of
attention \cite{amann}.

Most of the research on time delayed systems concentrates on deterministic systems. However
stochastic systems with time-delayed feedback are discussed as well, for instance, in the context of gene
regulation \cite{bratsun}. 

In this paper we investigate a simple discrete model with delay: a stochastic
process for a single binary variable which evolves according to its
own history. We show that this model can be mapped onto a
two-dimensional stochastic cellular automaton in such a way that the time-delayed
couplings become local. As a result, the autocorrelation function of the corresponding time series is related
to the critical properties of directed percolation. Our results are however more general and may be applied
to a large variety of systems.

\section{Reordering of the time series}

We consider a time series $x_t$ with a discrete time variable $t\in \mathbb{N}$ that evolves by non-local stochastic updates $(x_{t-n},x_{t-k}) \to x_t$ in such a way that the probability for the outcome $x_t$ is given by
\begin{equation}
\label{eq:updatingrule}
P(x_t) = f(x_t, x_{t-n},x_{t-k})\,,  \qquad t>n>k
\end{equation}
where $n$ and $k$ are two different delays. The initial configuration may be given by specifying $n$ subsequent elements of the time series, e.g. $x_0,\ldots,x_{n-1}$. The type of data represented by $x_t$ as well as the function $f$ is not restricted in any way; the only important ingredient is that a new entry of the time series $x_t$ depends on the previous values at times $t-n$ and $t-k$.

Here we show that irrespective of the structure of $x_t$ and $f$, it is possible to rearrange the time series in such a way that the couplings become local in a two-dimensional representation. More specifically, we show that it is possible to define a reordered series $y_t$ that evolves by updates $(y_{t-n},y_{t-s}) \to y_t$ with 
\begin{equation}
P(y_t) = f(y_t,y_{t-n},\,y_{t-s})\;,
\end{equation}
where $s$ is a different delay that switches between the values $1$ or $n+1$. The only condition for this transformation to work is that the original delays~$n$ and~$k$ have to be coprime, i.e., they have no common divisor other than~$1$. The reordered time series $y_t$ is related to the original one by 
\begin{equation}
x_t \;=\; y_{T(t)}
\end{equation} 
where the map $T(t)$ is given by
\begin{equation}
\label{eq:hoye}
T(t) \;=\; (tq) \mod n \,+\, n \lfloor t \rfloor_n\,.
\end{equation}
In this equation $\lfloor t \rfloor_n$ denotes integer division by $n$ while $q \in \{0,\ldots,n-1\}$ is an integer such that
\begin{equation}
(kq) \mod n=1\,.
\end{equation} 
Note that the existence of $q$ requires $n$ and $k$ to be coprime. The condition of coprimality of $n$ and $k$ guarantees that the mapping is one-to-one with the inverse
\begin{equation}
\label{eq:hoyeinverse}
T^{-1}(t) \;=\; (tk) \mod n \,+\, n \lfloor t \rfloor_n\,.
\end{equation}
Coprimality should not be seen as a shortcoming of the mapping since the number of coprimes to a given $n$, as given by the Euler's totient function $\phi (n)$, is known to increase sufficiently rapidly (faster than $\sqrt{n}$) so that in the large $n$ limit the transformation can be applied to systems with various delays $k$. 

To understand this transformation, the time line should be thought of as being divided into equidistant blocks of size $n$. The transformation simply reorders the time indices in each block. This reordering takes place in the first term on the r.h.s. of Eq.~(\ref{eq:hoye}) while the second term simply enumerates the blocks in such a way that they do not mix. 

\begin{figure}[t]
\centering\includegraphics[width=85mm]{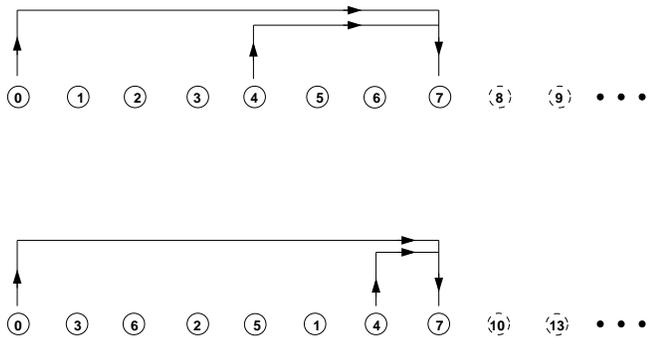}
\caption{\label{fig:visual}\small
Updating scheme in systems with delays $n=7$ and $k=3$ before (upper figure) and after (lower figure)
the transformation~(\ref{eq:hoye}). The arrows indicate the couplings needed to update the variable $x_7$ as a function of $x_0$ and $x_4$.}
\end{figure}

As an example let us consider the special case $n=7$ and $k=3$, hence $q=5$. One starts out with a given configuration $\{ x_0,x_1,\ldots,x_6\}$. From this initial condition the whole time series can be constructed by iteration of Eq.~(\ref{eq:updatingrule}). Applying the transformation Eq.~(\ref{eq:hoye}) the first $n$ sites are reordered by
\begin{equation}
\label{eq:example}
0\rightarrow 0\; ; \;
1\rightarrow 3\; ; \;
2\rightarrow 6\; ; \;
3\rightarrow 2\; ; \;
4\rightarrow 5\; ; \;
5\rightarrow 1\; ; \;
6\rightarrow 4.
\end{equation}
The same reordering scheme takes place in the subsequent blocks. As illustrated in Fig.~\ref{fig:visual}, this transformation preserves the long delay $n$ while the short delay $k$ is mapped onto $1$ or $n+1$. In the appendix we prove that this is also true in the general case as long as $n$ and $k$ are coprime.

\section{Two-dimensional representation}

\begin{figure}[t]
\centering\includegraphics[width=60mm]{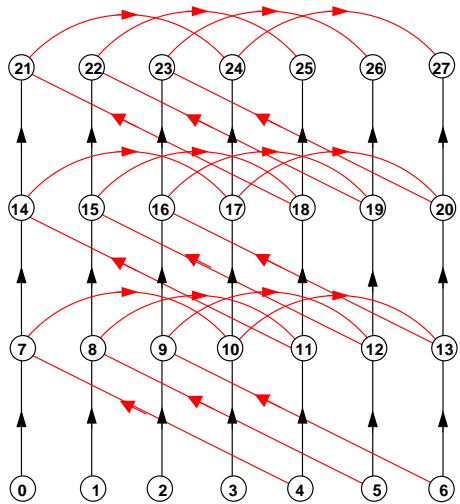}
\caption{\small\label{fig:original} (Color online)
Original time series $x_t$ with delays $n=7$ and $k=3$ in a two-dimensional representation~\cite{Giacomelli}. The time series is divided into segments of size $n$ and plotted line by line on top of each other. Starting with a given configuration $\{ x_0,x_1,\ldots,x_6\}$ the whole time series can be constructed by iteration of Eq.~(\ref{eq:updatingrule}), where the black arrows represent the long delay between $x_{t-n}$ and $x_t$ while the red ones indicate the short delay between $x_{t-k}$ and $x_t$. The figure is equivalent to the upper part of Fig.~\ref{fig:visual}, the only difference being that blocks of size $7$ are arranged as a two-dimensional grid. }
\end{figure}

\begin{figure}
\centering\includegraphics[width=85mm]{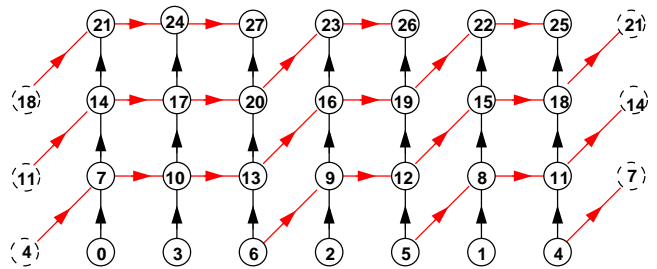}
\caption{\small\label{fig:transformedlattice}
(Color online)
Corresponding reordered time series in a two-dimensional representation according to Eq.~(\ref{eq:hoye}), where the bulk interactions become local.}
\end{figure}

The main purpose of the mapping described above is to arrange the reordered time series in a 1+1-dimensional plane in such a way that the interactions become local. Following Giacomelli and Politi~\cite{Giacomelli} the time series is divided into equidistant segments of $n$ elements which are arranged line by line on top of each other. This means that the index $t$ is mapped to the position
\begin{equation}
\left(\begin{array}{c}
 x\\y \end{array}\right)\;=\;
\left(  \begin{array}{c}t \mod n \\ \lfloor t \rfloor_n 
\end{array}\right)
\end{equation}
Obviously this procedure turns the long delay~$n$ into a local interaction. For example, drawing the original time series $x_t$ in such a 1+1-dimensional representation the long delay $n$ turns into a nearest-neighbor interaction in vertical direction while the short delay is still non-local (see Fig.~\ref{fig:original}). However, drawing the reordered time series $y_t$ in a 1+1-dimensional representation the long as well as the short delay become local, as shown in Fig.~\ref{fig:transformedlattice}. Note that the corresponding boundary conditions are not periodic but shifted in vertical direction, connecting subsequent blocks periodically in a spiral-like manner. 
 
\begin{figure}
\centering\includegraphics[width=75mm]{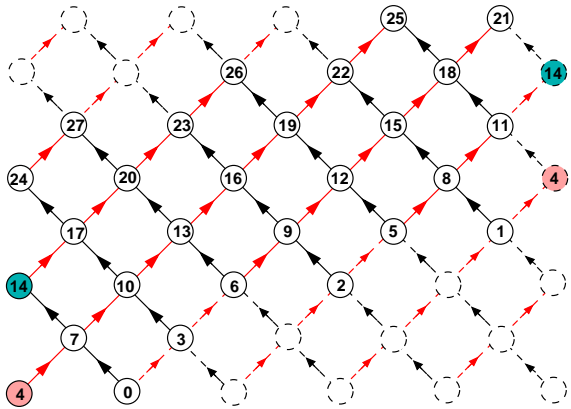}
\caption{\small\label{fig:distortedlattice}
(Color online)
The same lattice as in Fig.~\ref{fig:transformedlattice} plotted in a way such that all updates in the bulk are geometrically identical. The boundary conditions one obtains can be seen as
being time-delayed (time flows upwards).}
\end{figure}

\begin{figure}
\begin{center}
\vglue 10mm
\includegraphics[width=65mm]{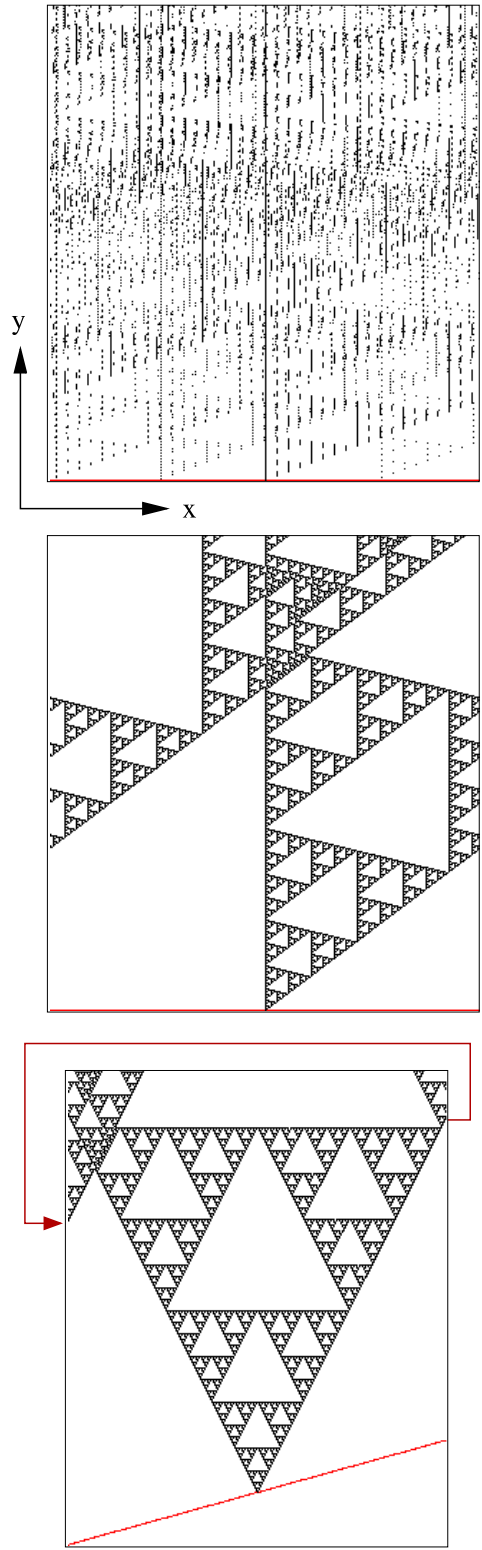}
\caption{
\label{fig:sp}
Visual representation of the time series generated by the update rule~(\ref{eq:sp}). The three panels correspond to the examples shown above in Figs.~\ref{fig:original},~\ref{fig:transformedlattice}, and \ref{fig:distortedlattice} (see text).
}
\end{center}
\end{figure}

Although the couplings in Fig.~\ref{fig:transformedlattice} are local, they are still biased towards north-east. Moreover, the coupling scheme exhibits vertical dislocation lines. As illustrated in Fig.~\ref{fig:distortedlattice}, these irregularities can be removed by redrawing the figure in such a way that all updates have the same orientation in the $xy$-plane. This representation allows one to relate the original time series with 1+1-dimensional cellular automata on a  tilted square lattice. However, by rearranging the lattice one obtains skewed boundary conditions with a \textit{non-local delay} connecting blocks with vertical distance $n-k$.
 
To demonstrate how the transformation works let us consider a simple deterministic update rule where $x_t \in {0,1}$ is a binary time series. The update rule is given by the Boolean function
\begin{equation}
\label{eq:sp}
x_t \;=\; x_{t-n} \oplus x_{t-k} \;=\;
\left\{
\begin{array}{ll}
1 & \mbox{ if } x_{t-n} \neq x_{t-k} \\
0 & \mbox{ otherwise } \\
\end{array}
\right.\,,
\end{equation}
where $\oplus$ denotes a logical XOR. Such a binary time series can be visualized by plotting $x_t$ as black and white pixels at position $(x,y)$. The results are shown in Fig.~\ref{fig:sp}, where we used the delays $n=300$ and $k=227$. Starting with a single non-zero entry in the initial state $x_{n/2}=1$ the iteration of the update rule (\ref{eq:updatingrule}) produces an irregular pattern of pixels, which is shown in the upper panel of Fig.~\ref{fig:sp}. Applying the transformation~(\ref{eq:hoye}) the pixels are ordered with a bias to the north-east, resulting in a tilted Sierpinsky gasket (see middle panel). Finally, plotting the same data in such a way that the tilt is removed one obtains the usual form of the Sierpinsky gasket (see lower panel). However, as exemplified by the arrow, the boundary conditions are no longer periodic, instead they involve a non-local shift $\delta=n-k$ in vertical direction.  

\section{Stochastic update rule related to directed percolation}

\begin{figure*}[t]
\centering\includegraphics[width=150mm]{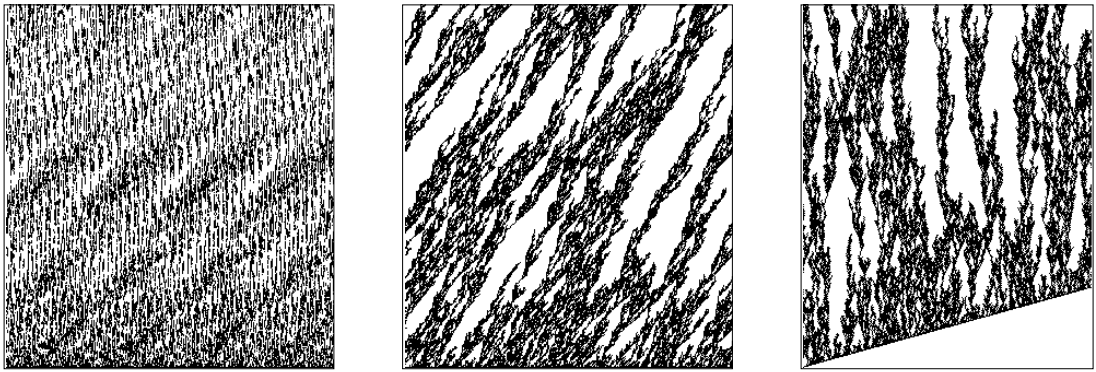}
\caption{\label{fig:dp}\small
Analogous representation of a stochastic time series with $n=300$ and $k=227$ evolving by a directed percolation update rule at the critical point $p=p_c$. Here the iteration starts with a sequence $111\ldots $\,.}
\end{figure*}

Let us now turn to a simple but non-trivial example of a \textit{stochastic} update rule. For a binary time series a probabilistic update $(x_{t-n},x_{t-k}) \to x_t$ according to Eq.~(\ref{eq:updatingrule}) is determined by 
\begin{eqnarray}
\label{eq:Example}
f(1,x_i,x_j) &=& p_{x_i,x_j} \\
f(0,x_i,x_j) &=& 1-p_{x_i,x_j} \nonumber 
\end{eqnarray}
with four control paramters $p_{00},p_{01},p_{10},p_{11}$. In what follows let us assume that $p_{00}=0$. In this case the time series consisting of zeroes is a fixed point of the dynamics. In non-equilibrium statistical physics such a configuration which can be reached but not be left is called an absorbing state. Whether or not this absorbing state is stable against perturbations depends on the magnitude of the remaining control parameter $p_{01},p_{10}$ and $p_{11}$. For example, setting
\begin{equation}
p_{01}=p_{10}=p\,,\qquad p_{11}=2p-p^2\,
\end{equation}
and varying $p$ between $0$ and $1$ one observes the following phenomenological behavior: 
\begin{itemize}
\item If $p$ is very small the time series quickly approaches the absorbing series consisting of zeroes. 
\item For large $p$ the dynamics approaches a fluctuating steady state with a non-vanshing stationary expectation value of~$x_t$. The probability to reach the absorbing configuration is very low and decreases with increasing $n$. 
\item At a certain threshold $p_c  \approx 0.6447$ one observes a power-law decay of the density in a finite temporal range which grows with $n$. 
\end{itemize}
This behavior reminds of a non-equilibrium phase transition from a fluctuating phase into an absorbing state. In fact, using the transformation~(\ref{eq:hoye}) the update rule becomes equivalent to that of a Domany-Kinzel cellular automaton~\cite{DomanyKinzel}, which is known to exhibit a second-order phase transition belonging to the universality class of Directed Percolation (DP)~\cite{Kinzel,Hinrichsen,Odor,Lubeck}. Using the present notation the 1+1-dimensional Domany-Kinzel model is defined on a tilted square lattice with coordinates $(x,y)$. Each lattice site can be either active ($s(x,y)=1$) or inactive ($s(x,y)=0$). The model evolves by parallel updates, i.e. the new horizontal line at $y+1$ is obtained by setting
\begin{equation}
s(x,y+1) :=\left\{ 
\begin{array}{ll}
1 & \mbox{ with probability } p_{ s(x-1,y),\,s(x+1,y)}\\
0 & \mbox{ otherwise. }
\end{array}\right.
\end{equation}
For the choice $p_{01}=p_{10}=p$ and $p_{11}=2p-p^2$ the Domany-Kinzel model reduces to directed bond percolation. This model is known to exhibit a continuous phase transition belonging to the universality class of directed percolation at the critical point $p_c = 0.6447001(2)$ if the system size is infinite. In fact, as shown in Fig.~\ref{fig:dp}, the transformation~(\ref{eq:hoye}) maps an apparently disordered time series into an ordered one, where typical DP cluster can be seen. 

It should be stressed that in the present model the corresponding DP process takes place on a \textit{finite} lattice so that for any finite $n$ there is no phase transition in a strict sense. Nevertheless it is possible to observe the typical signatures of DP critical behavior within a certain temporal range which grows with $n$, as will be shown in the following.

\subsection{Two-point correlation function}

\begin{figure}
\centering\includegraphics[width=85mm]{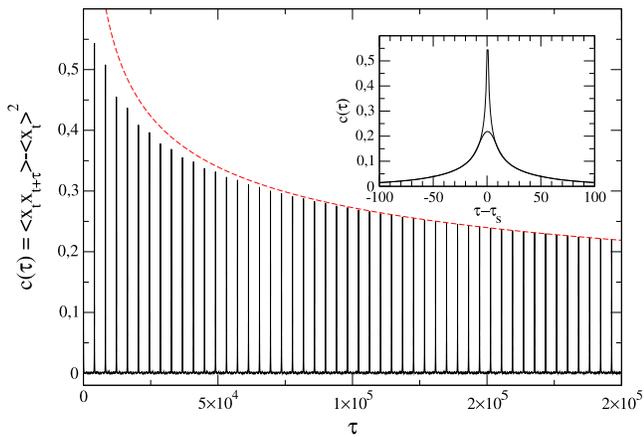}
\caption{\label{fig:dpcorr}\small
Two-point correlation function of the time series $c(r)$ defined in Eq.~(\ref{eq:twpoint}). The dashed line demonstrates that the spike peaks decay as $\tau_s^{-2\beta/\nu_\parallel}$. The inset shows a zoom of the first and the last spike.}
\end{figure}

In order to see a signature of DP critical behavior we tried to identify the critical exponents in 1+1 dimensions
\begin{equation}
\beta=0.276486(8),\;\; \nu_\perp=1.098654(4),\;\; \nu_\parallel=1.733847(6).
\end{equation}
To this end we iterated the time series slightly above the critical point $p=p_c+0.001=0.6448$ and measured the connected part of the two-point correlation function 
\begin{equation}
\label{eq:twpoint}
c(\tau) = \langle x_t x_{t-\tau}\rangle - \langle x_t \rangle^2 \,.
\end{equation}
Before the average was taken the time series was equilibrated over $2 \times 10^9$ iterations in order to reach a stationary state. Moreover, we chose a very large delay $n=4096$ to prevent the system from entering the zero sequence due to finite-size effects. Figure ~\ref{fig:dpcorr} shows a train of spikes at regularly spaced times $\tau_s$ which are caused by correlations between subsequent rows in the two-dimensional representation. The asymptotic envelope of these spikes seems to obey an asymptotic power law 
\begin{equation}
c(\tau_s) \sim \tau_s^{-\gamma }
\end{equation}
with an exponent $\gamma = 0.32(4)$. The relation to DP predicts this exponent to be given by
\begin{equation}
\gamma = 2 \beta / \nu_\parallel \approx 0.318.
\end{equation}
A similar attempt to obtain the spatial correlation exponent $2 \beta /\nu_\perp$ from the form of a single spike (see inset) fails. This can be explained as follows. In the central panel of Fig.~\ref{fig:dp} the form of the spike would correspond to a correlation function in horizontal direction, whereas in the symmetrized representation shown in the right panel this correlation function would be tilted. Therefore, the spike profile is given by an interplay of both exponents $\nu_\parallel$ and $\nu_\perp$, making it difficult to distinguish between them.

\subsection{Dynamical scaling of the pair connectedness function}

\begin{figure}
\centering\includegraphics[width=85mm]{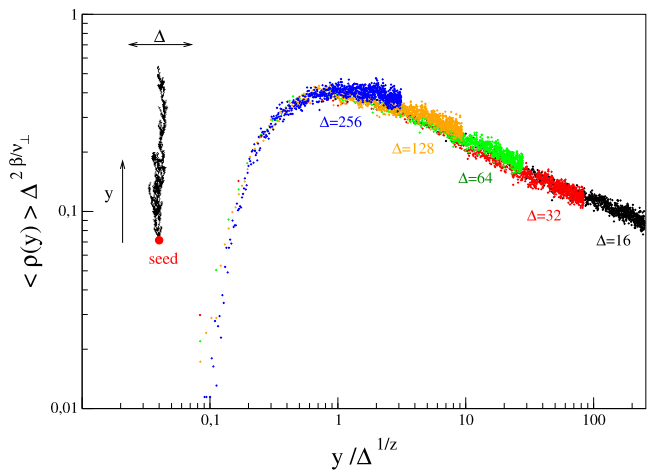}
\caption{\label{fig:dpcollapse}\small (Color online)
Data collapse for the counterpart of the pair connectedness function for $n=512$ and $k=1$.}
\end{figure}

In order to identify the DP critical exponents more clearly, we measured the counterpart of the so-called pair connectedness function~(see e.g.~\cite{Hinrichsen}) at criticality $p=p_c$. The iteration starts with a single active seed $x_n=\delta_{n,n_0}$. After reordering the time series and representing it in the right panel of Fig.~\ref{fig:dp}, we measure the density $\rho(\Delta,y)$ in vertical distance $y$ and horizontal distance $\Delta$ from the seed. Since the pair connectedness function is known to obey the scaling form
\begin{equation}
\rho(\Delta,y) = \Delta^{-2\beta/\nu_\perp} \, h(y/\Delta^{1/z})
\end{equation}
with a universal scaling function $h$, the exponents can be determined by plotting $\rho(\Delta,y)\Delta^{2\beta/\nu_\perp}$ versus $y/\Delta^{1/z}$ in such a way that data sets for different values of $\Delta$ collapse. Plugging in the known exponents of DP one obtains a convincing data collapse, as shown in Fig.~\ref{fig:dpcollapse}. This confirms unambiguously the critical behavior we expected to get.

\section{Discussion}
\label{Discussion}

In this paper we have shown how a time series evolving according to a nonlocal
update rule can be mapped onto a local process in two dimensions with special
time-delayed boundary conditions. One interesting question is whether this
result holds also for the continuous case. For example, discretizing a 
differential equation of the form
\begin{equation}
\label{eq:continuum}
\dot{x}(t) = F[x(t)] + G[x(t-\tau)]
\end{equation}
with delay $\tau$ and arbitrary functions $F$ and $G$ one gets
\begin{equation}
x_m - x_{m-1} = h [F(x_{m})+G(x_{m-k})]\,,
\end{equation}
where $h$ is the step size and $k=\tau/h$ is the discrete analogue of the delay. In the present paper we have shown that the equation 
\begin{equation}
\label{eq:final}
x_m - x_{m-s} = h [F(x_{m})+G(x_{m-k})]
\end{equation}
with $1<k<s$ exhibits (up to boundary conditions) the same properties as long as $s$ and $k$ are coprime. The question is whether
it is possible to find an appropriate limit of (\ref{eq:final}) and recover (\ref{eq:continuum}). Work in this direction
is currently under way~\cite{DH}. 

\vspace{5mm}
\noindent
Acknowledgements:
This work was motivated by a diploma thesis of M. Ackermann, who studied a delayed time series with an update rule related to directed percolation. S. R. Dahmen would like to thank the Alexander von Humboldt Foundation for financial support and the University of W\"urzburg for the hospitality.

\appendix
\section{Proof of the transformation}
\label{proof}

The transformation~(\ref{eq:hoye}) is proven in two steps. First we show that two sites of the time series separated by a time delay of $\Delta t = n$ are still separated by the same delay after the transformation. Then we show that sites that were originally separated by a time delay $\Delta t = k$ are mapped onto new sites either with a time delay $\Delta t = 1$ or $\Delta t = n+1$.

We start with the sites separated by $n$ time steps, that is $t$ and $t-n$. According to the transformation
rule we get 
\begin{eqnarray}
T(t)-T(t-n)&=& (tq)\mod n +n \lfloor t\rfloor_n \nonumber \\
&&- \underbrace{(t-n)q\mod n}_{ = (tq)\;mod\; n} -n\lfloor (t-n) \rfloor_n \nonumber\\
&=& n \left( \lfloor t\rfloor_n -\lfloor (t-n)\rfloor_n\right)
\;=\;n\,,
\end{eqnarray}
where we used the fact that for two sites of different blocks (which is always the case here) one has $\lfloor t\rfloor_n -\lfloor (t-n)\rfloor_n = 1$.


Next we consider the case where sites have a delay of~$k$, i.e.
\begin{eqnarray}
\label{delay1}
T(t)-T(t-k) &=& (tq)\mod n -\left( (t-k)q\right)\mod n \nonumber \\
&&\,+\, n\left(\lfloor t\rfloor_n - \lfloor t-1\rfloor_n\right)
\end{eqnarray}
Here we have to distinguish two cases. If $t$ is a multiple of $n$ this expression reduces to
\begin{eqnarray}
T(t)-T(t-k) &=& -\underbrace{(-kq)\mod n}_{=n-1} + n(\underbrace{\lfloor t\rfloor_n - \lfloor t-1\rfloor_n}_{=1}) \nonumber \\
&=& 1
\end{eqnarray}
On the other hand, if $t$ is not a multiple of $n$ we get
\begin{widetext}
\begin{eqnarray}
T(t)-T(t-k) &=& (tq)\mod n - (tq-kq)\mod n \nonumber
 +  n\left(\lfloor t\rfloor_n - \lfloor t-1\rfloor_n\right)\nonumber \\
&=& \underbrace{(tq)\mod n}_{\in\{1,...,n-1\}} 
 - \Bigl(\underbrace{(tq)\mod n}_{\in\{1,...,n-1\}}- \underbrace{(kq)\mod n}_{=1}\Bigr)\mod n \nonumber 
         +   n\left(\lfloor t\rfloor_n - \lfloor t-1\rfloor_n\right)\nonumber \\
&=& 1 \,+\,  n\left(\lfloor t\rfloor_n - \lfloor t-1\rfloor_n\right) 
\;=\; \left\{\begin{array}{ll} 1   & \mbox{ if } \hspace{3mm} t \mod n \geq k \\ 
 		  n+1 & \mbox{ if } \hspace{3mm} t \mod n < k \end{array} \right.
\end{eqnarray}

\end{widetext}

\end{document}